\documentclass[preprint,12pt, a4paper]{elsarticle}

\usepackage{overpic}
\usepackage{amssymb}
\usepackage{hyperref}
\usepackage{xcolor}

\journal{XXX}

\begin{document}

\begin{frontmatter}



\title{AutoCT: Automated CT registration, segmentation, and quantification}


\author[1]{Zhe Bai\corref{cor1}}
\author[2]{Abdelilah Essiari}
\author[2]{Talita Perciano\corref{cor1}}
\author[2,3,4]{Kristofer E. Bouchard}
\address[1]{Applied Mathematics and Computational Research Division, Lawrence Berkeley National Laboratory}
\address[2]{Scientific Data Division, Lawrence Berkeley National Laboratory}
\address[3]{Biological Systems and Engineering Division, Lawrence Berkeley National Laboratory}
\address[4]{Helen Wills Neuroscience Institute and Redwood Center for Theoretical Neuroscience, UC Berkeley}

\cortext[cor1]{Corresponding authors: All correspondence should be sent to both zhebai@lbl.gov and tperciano@lbl.gov.}

\begin{abstract}
The processing and analysis of computed tomography (CT) imaging is important for both basic scientific development and clinical applications. In AutoCT, we provide a comprehensive pipeline that integrates an end-to-end automatic preprocessing, registration, segmentation, and quantitative analysis of 3D CT scans.  The engineered pipeline enables atlas-based CT segmentation and quantification leveraging diffeomorphic transformations through efficient forward and inverse mappings. The extracted localized features from the deformation field allow for downstream statistical learning that may facilitate medical diagnostics. On a lightweight and portable software platform, AutoCT provides a new toolkit for the CT imaging community to underpin the deployment of artificial intelligence-driven applications.

\end{abstract}


\end{frontmatter}


\noindent
\section{Motivation and significance}

Computed tomography (CT) stands as one of the most prevalent medical imaging modality in the world. Nevertheless, interpreting and analyzing CT data demands substantial professional expertise and efforts, posing challenges for acute diagnoses and prognoses. 
In contrast to MRI, CT images typically exhibit lower spatial resolution and contrast levels accompanied by systematic noise. This combination presents a formidable challenge for even proficient radiologists tasked with analyzing localized diseases or lesions in specific regions in limited time. 

The registration and segmentation of CT images constitute essential processes in medical image analysis~\cite{antonelli2022medical}. Registration enables the integration of features across multiple scans, while segmentation delineates structures of interest, allowing for quantitative analysis. Current image registration algorithms utilize deformable mappings based on elastic, fluid-based deformations that improve upon linear or rigid transformations.  
The Demons~\cite{thirion1998image, vercauteren2009diffeomorphic} algorithm pioneered diffeomorphic deformations, effectively preserving anatomical structures. Nonlinear registration algorithms, such as those provided by Advanced Normalization Tools (ANTs)~\cite{avants2010optimal, avants2011reproducible}, deliver robust and adaptable solutions for complex anatomical variations. 
Recently, deep learning approaches especially convolutional neural networks (CNNs)~\cite{litjens2017survey, kuppala2020overview}, have demonstrated success in deformable image registration, while hybrid models~\cite{fu2020deep, chen2021deep} that integrate CNNs with classical methods, aim to balance precision with computational efficiency.

Here, we introduce the AutoCT software, which provides an integrated and automated pipeline for individual 3D CT scans that enables robust and efficient image analysis. This pipeline potentially reduces the human cost required for medical diagnosis and prognosis of traditional approaches. By streamlining preprocessing tasks, AutoCT enables medical professionals to devote more time to interpreting results and making informed clinical decisions. Moreover, the software automates the identification and analysis of anatomical structures, potentially enabling precise localization of abnormalities.

\begin{figure}
\centering
\includegraphics[width=1\textwidth]{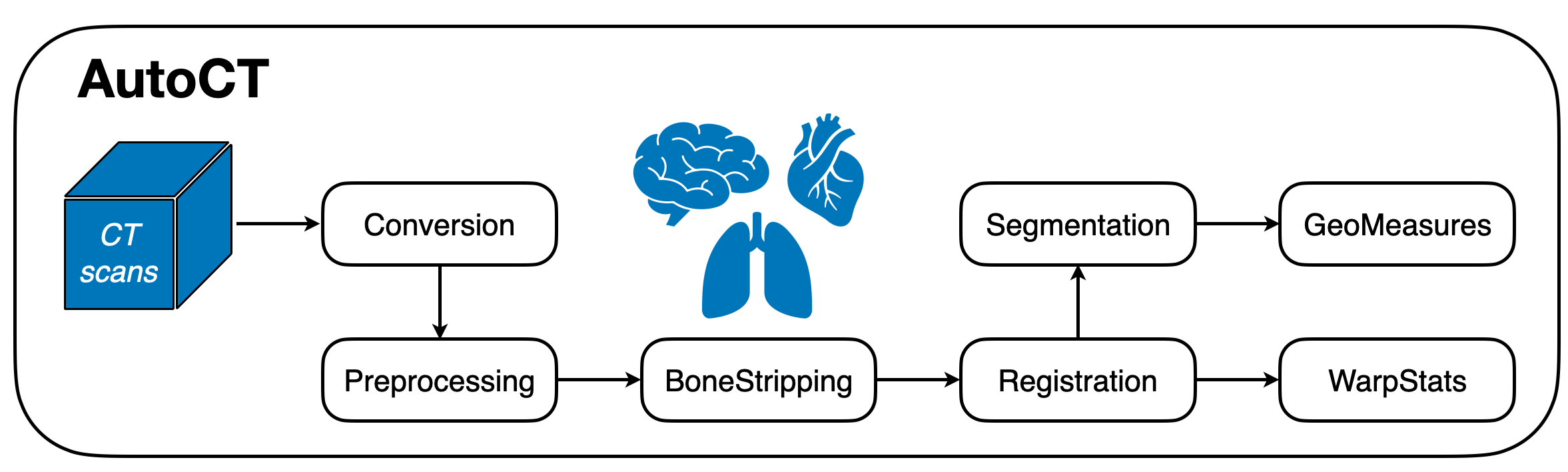}
\caption{Workflow of AutoCT: a comprehensive pipeline including CT scans conversion, preprocessing, bone stripping, registration, segmentation and quantification.}
\label{workflow}
\end{figure}

\section{Software description}
AutoCT is written in Python 3.7, making use of few external packages (principally \emph{dcm2niix}~\cite{li2016first}, ~\emph{FSL}~\cite{jenkinson2012fsl}, ~\emph{ANTs}~\cite{avants2010optimal}). 
This end-to-end pipeline integrates automatic conversion, preprocessing, bone striping, registration, segmentation, and quantitative analysis for 3D CT scans.  
\subsection{Software architecture}
The overall architecture, as shown in Figure~\ref{workflow} enables a user to input original DICOM images, which undergo a comprehensive pipeline including file conversion, preprocessing, bone stripping, registration, and segmentation, and outputs quantitative information based on a user-defined anatomical template and atlas. Therefore, users have the capability to acquire extracted features, including warp (i.e., deformation) statistics with respect to the template and segmented geometric measurements from each individual region of interest in the anatomical structure for subsequent analysis.

\subsection{Software functionalities}
\begin{figure}
    \centering
    \includegraphics[width=1\textwidth]{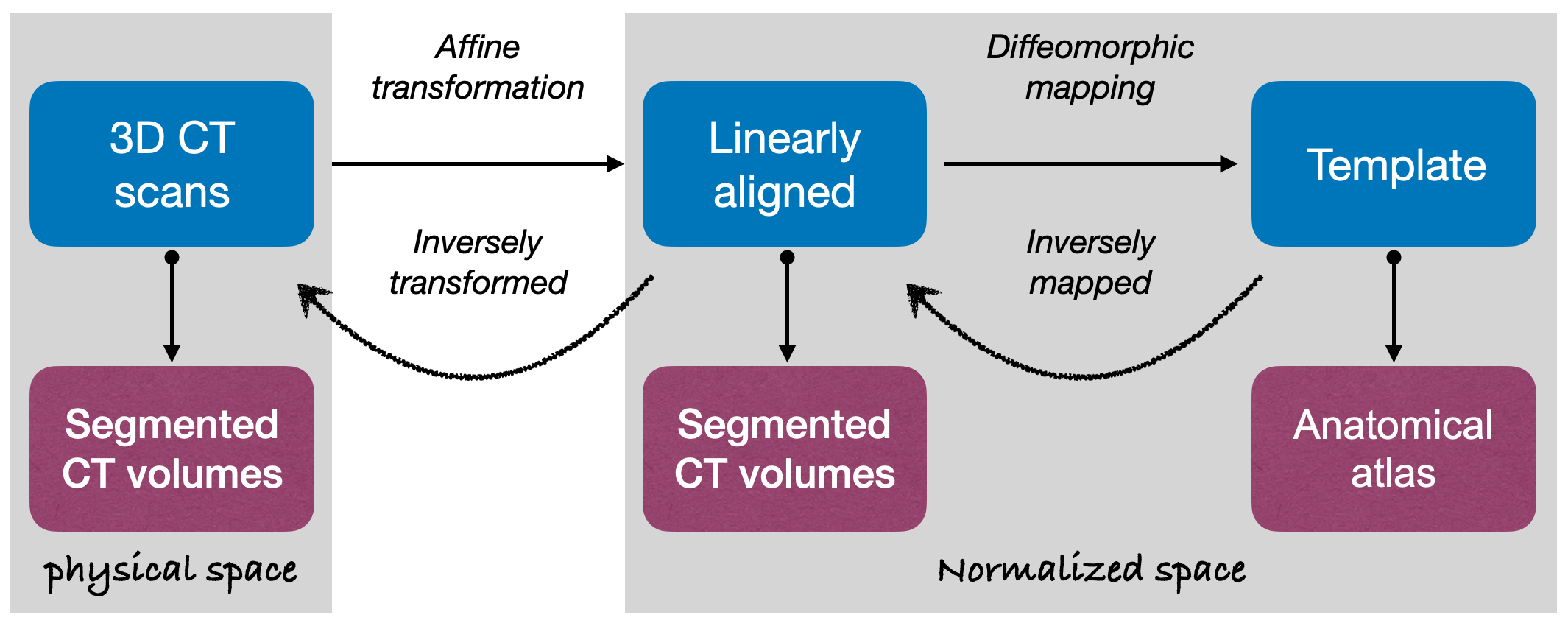}
    \caption{Processing of joint image registration and segmentation, based on a user-specified template and anatomical atlas.}
    \label{method}
\end{figure}

Figure~\ref{workflow} summarizes the workflow of AutoCT connecting seven modules. First, the \emph{Conversion} module takes in each subject's raw CT scans as DICOM images and converts them to a 3D image volume in NIfTI format. The \emph{Preprocessing} module takes in NIfTI images, standardizes the image orientation, resamples the volume to a standardized voxel size, performs bias correction~\cite{tustison2010n4itk}, and then pre-aligns the 3D volume to a canonical, standardized space, such as MNI~\cite{brett2002problem} coordinates. Following preprocessing, the \emph{BoneStrip} module extracts the soft tissue of interest from surrounding bone using binary mask created through successive procedures of intensity thresholding, hole filling and Gaussian smoothing~\cite{woolrich2009bayesian}. These bone-stripped CT scans are then registered to a user-defined template in the \emph{Registration} module. The \emph{Registration} module deploys a diffeomorphic mapping built on Advanced Normalization Tools~\cite{avants2010optimal, avants2011reproducible} and performs a smooth and invertible transformation between the CT volumes and the reference. Using a diffeomorphism ensures a point in the physical domain is mapped to a corresponding point in the standardized or normalized domain, and the mapping is characterized by its differentiable and invertable properties\cite{leslie1967differential, krantz2002implicit}. This process enables complex spatial deformations while preserving the volumes topology. 
Figure~\ref{method} delineates the joint image registration and segmentation process. By applying the inverse diffeomorphic mapping operator (obtained from the registration stage) to the desired anatomical atlas, AutoCT parcellates the bone-stripped CT volumes in the normalized space, and then uses the inverse affine transformation to realize the segmentation in the physical space in the \emph{Segmentation} module. Finally, the deformation field is used to generate statistical information for the nonlinear transformation, including the mean, standard deviation and entropy of the Jacobian of the mapping~\cite{andersson2007non} for the warped image and is exported to the \emph{WarpStats} module. In parallel, the \emph{GeoMeasures} module extracts geometric measurements for each segment, for example the volume and surface area for characterization in both the physical and normalized spaces.  

This scientific software has been packaged and tested using Docker container technology. The AutoCT docker image contains all the required dependencies and libraries needed to run the workflow.
Users can easily reproduce the results of the pre-packaged \emph{illustration} workflow, and execute customized workflows using their own data and share the results with others using the docker archiving facilities. 
The AutoCT docker container provides the option to run the workflow using the command line or the interactive python-based Jupyter notebooks. Additionally, an interactive graphical interface, based on Jupyter \emph{Widgets}, is available to assist users in executing the different stages of the workflow. 
Finally, the docker image is used as part of continuous integration in the development process. When new modifications to AutoCT codes are checked into the version control system, a docker image will be automatically built and a container will be launched to run the pipeline and compare to expected results for evaluation.

\section{Illustrative Examples}
\begin{figure}
    \centering
    \begin{overpic}[width=0.63\textwidth]{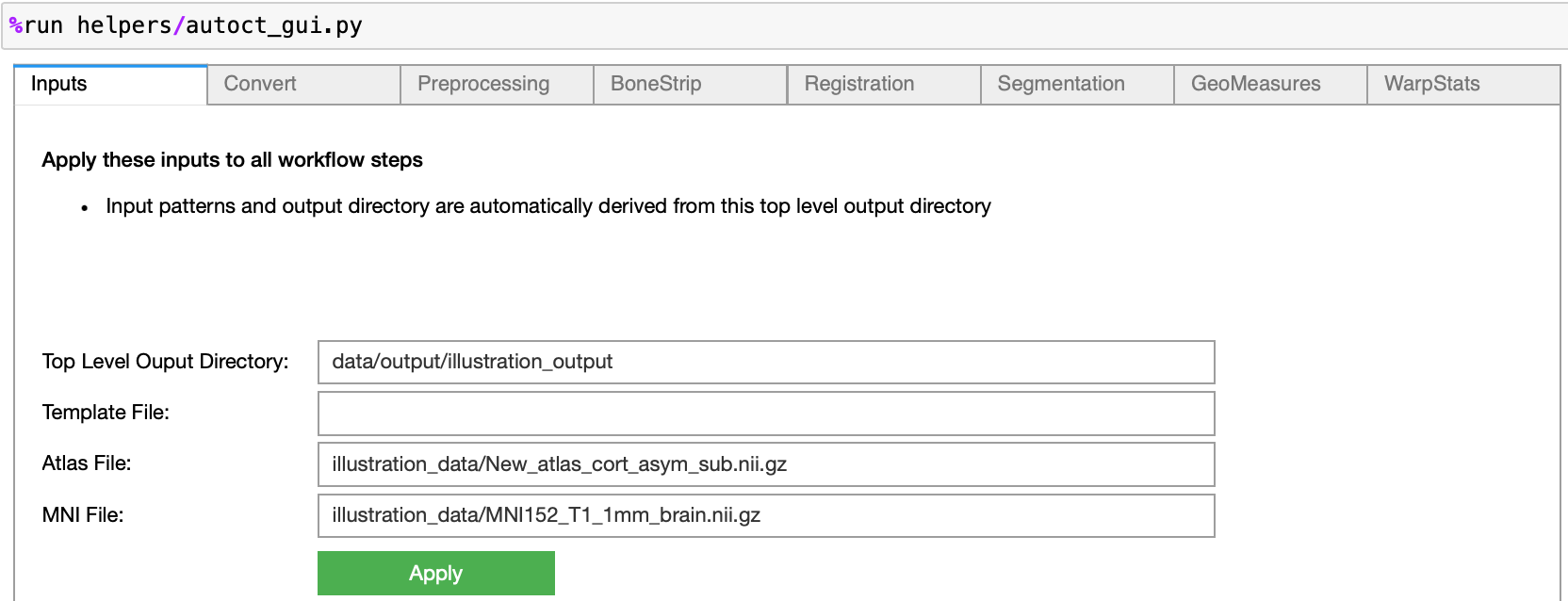}
    \put(-6,36.5){\footnotesize(a)}
    \end{overpic}
    \begin{overpic}[width=0.3\textwidth]{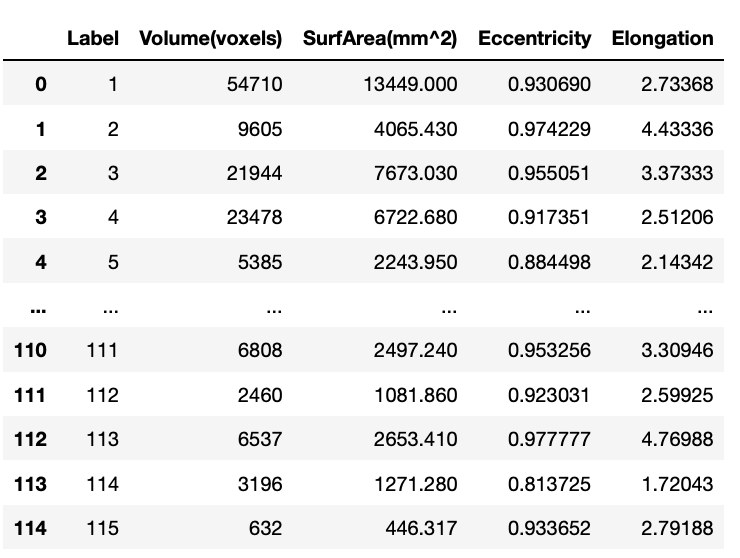}
    \put(-3,76){\footnotesize(b)}
    \end{overpic}\\
    \begin{overpic}[width=0.496\textwidth]{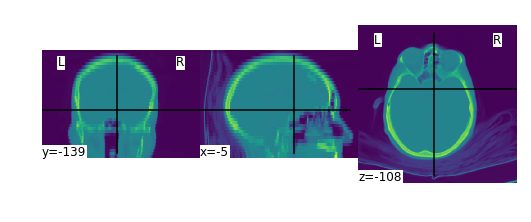}
    \put(-1,34.5){\footnotesize(c)}
    \end{overpic}
    \begin{overpic}[width=0.495\textwidth]
    {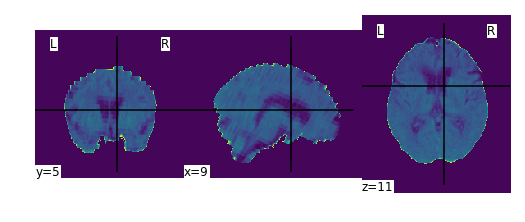}
    \put(-2,35){\footnotesize(d)}
    \end{overpic}\\
    \begin{overpic}[width=0.495\textwidth]{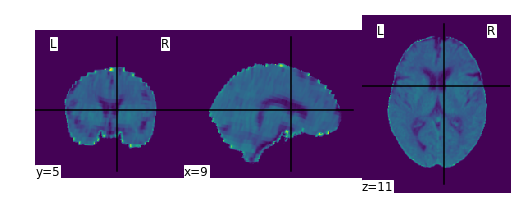}
    \put(-1.5,36){\footnotesize(e)}
    \end{overpic}
    \begin{overpic}[width=0.495\textwidth]{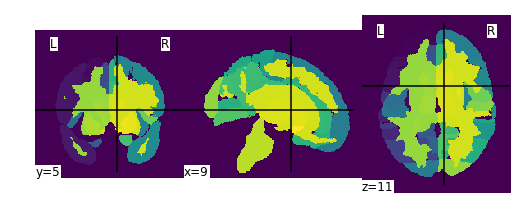}
    \put(-2,36){\footnotesize(f)}
    \end{overpic}
    \caption{Illustrative example: (a) interactive GUI design (b) output measurements (c) original CT sample
    (d) preprocessed, bone-stripped CT sample (e) warped image after registration (f) segmented CT volume in the physical space.
    }
    \label{example}
\end{figure}

In this section, we provide an illustrative example of the AutoCT application by running it on a publicly available source of CT scans~\cite{kaggle19}. The configuration options for the input files are shown in Figure~\ref{example}(a). For this illustration, we use a standard MNI Template of T1-weighted MRI~\cite{brett2002problem} as the template for pre-alignment, and a combined Harvard-Oxford cortical and subcortical structural atlases~\cite{makris2006decreased} for image segmentation. Figure~\ref{example}(c-f) present the original, skull-stripped, and warped image after registration in the MNI space as well as the segmented CT volume in the subject's physical space. The output data shown in Figure~\ref{example}(b) highlights the geometric measures of $115$ regions covering cortical and subcortical areas of the brain, where each label represents one region obtained from the segmentation module. This example is illustrated in a Jupyter notebook within the open-source repository for reproducibility.   

\section{Impact}
AutoCT not only expedites the processing of CT scans of complex tissues, it also enhances the localized analysis accuracy and reliability of quantification and potential diagnostics. The user-friendly setup allows one to specify a template or atlas tailored to a particular input, such as brain or lung CT scans. The output extracted features can further be studied and connected to research in simulation or clinical results for predictive model design. The Docker containerization approach with a user-friendly GUI enables a diverse user group to improve reproducibility and collaborative development. Therefore, this portable software may empower radiology and medical imaging research and facilitate practitioners to deliver more efficient and rigorous CT image analysis. 

\section{Conclusion}
This paper introduces AutoCT, an open-sourced, Docker image-based software package designed for the processing, registration, segmentation and analysis of CT scans. This automated pipeline represents a significant advancement in the analysis of CT volumes that may enhance research and medical capabilities. Notably, AutoCT features a modular and flexible architecture, making it suitable for diverse applications in domains ranging from imaging science to medical research. Future developments will expand the software's functionality to incorporate additional insights gleaned from clinical studies, enabling comprehensive assessments of CT scans across various modalities. Additionally, sensitivity analysis of the extracted quantitative features will be performed to generate data-driven predictive models for diagnosis and prognosis, further enhancing its effectiveness in the realms of medical imaging research.

\section*{Declaration of competing interest}
The authors declare that they have no known competing financial interests or personal relationships that could have appeared to influence the work reported in this paper.

\section*{Acknowledgements}
We would like to thank Esther Yuh, Geoffrey Manley, and Wibe de Jong for valuable discussions.  We acknowledge support by the U.S. Department of Energy, Office of Science, under Award Number DE-AC02-05CH11231. ZB gratefully acknowledges support from the U.S. Department of Energy, Office of Science, SciDAC/Advanced Scientific Computing Research. This research used resources of the National Energy Research Scientific Computing Center (NERSC), a U.S. Department of Energy Office of Science User Facility located at Lawrence Berkeley National Laboratory, operated under Contract Number DE-AC02-05CH11231.

\bibliographystyle{elsarticle-num} 
\bibliography{references}

\section*{Required Metadata}
\label{}

\section*{Current code version}
\label{}


\begin{table}[!h]
\begin{tabular}{|l|p{6.5cm}|p{6.5cm}|}
\hline
\textbf{Nr.} & \textbf{Code metadata description} & \textbf{Please fill in this column} \\
\hline
C1 & Current software version & AutoCT 1.1.2 \\
\hline
C2 & Permanent link to code/repository used for this code version & \url{https://github.com/zhbai/AutoCT} \\
\hline
C3  & Permanent link to Reproducible Capsule & \\
\hline
C4 & Legal Code License   & BSD 3-Clause License \\
\hline
C5 & Code versioning system used & Git \\
\hline
C6 & Software code languages, tools, and services used & Python, Docker, Jupyter notebook \\
\hline
C7 & Compilation requirements, operating environments \& dependencies & Mac/Linux/Unix, 
python==3.7, ants==2.3.1, fsl==5.0.10, dcm2niix==1.0, nibabel==3.2.0, joblib==1.0.1, nipype==1.6.0, pytest==6.0.1, pandas==1.1.2, ipywidgets==7.5.1\\
\hline
C8 & If available Link to developer documentation/manual & \url{https://github.com/zhbai/AutoCT/#README} \\
\hline
C9 & Support email for questions & tperciano@lbl.gov, zhebai@lbl.gov \\
\hline
\end{tabular}
\caption{Code metadata (mandatory)}
\label{} 
\end{table}




\end{document}